\documentclass[aps,prb,twocolumn,floatfix,showpacs,groupedaddress]{revtex4}
\usepackage{amsmath,amsfonts,amssymb,graphicx,times,bbm,natbib}
\usepackage[T1]{fontenc}
\usepackage{times}
\usepackage{color}
\begin{document}

\title{Spin and charge dynamics of the one-dimensional extended Hubbard model}

\author{H. Benthien}
\affiliation{Fachbereich Physik, Philipps-Universit\"{a}t, 
D-35032 Marburg, Germany}

\author{E. Jeckelmann}
\affiliation{Institut f\"{u}r Theoretische Physik, Universit\"{a}t Hannover, 
D-30167 Hannover, Germany}

\date{\today}

\begin{abstract}
We investigate the dynamical spin and charge structure factors and
the one-particle 
spectral function of the one-dimensional extended Hubbard model at half
band-filling 
using the dynamical density-matrix renormalization group method. 
The influence of the model parameters on these
frequency- and momentum-resolved 
dynamical correlation functions 
is discussed in detail for the Mott-insulating regime.
We find quantitative agreement between our numerical 
results and experiments for the optical 
conductivity, resonant inelastic X-ray scattering, neutron scattering,
and angle-resolved photoemission spectroscopy in the quasi-one-dimensional
Mott insulator SrCuO$_2$.

\end{abstract}

\pacs{71.10.Fd, 71.10.Pm, 79.20.Ap, 78.20.Bh}

\maketitle

\section{Introduction}
The spectral properties of low-dimensional correlated electron systems have
attracted a significant amount of interest in recent years.~\cite{dionys} 
In particular, the
energy and momentum resolution of spectroscopic techniques have been greatly
improved. These methods now provide
plenty of information about the dispersion and intensity
of electronic excitations in low-dimensional materials. For instance, 
the dynamical separation of charge and spin degrees of freedom, which is
generic in the theory of quasi-one-dimensional correlated systems,
has been observed directly in angle resolved photoemission spectroscopy (ARPES)
of strongly anisotropic cuprate compounds.~\cite{kim,koi06} 
From the viewpoint of theory, however, 
calculating the response functions that are probed in these experiments 
remains a great challenge 
and thus a coherent quantitative description of these materials is still
lacking.

Few exact analytical results are available for dynamical response functions
in itinerant correlated electron systems.
In gapless systems dynamical correlation functions can be evaluated
analytically with field-theoretical methods in the limit of small excitation
energies.~\cite{voit} 
Gapped systems can also be treated
analytically in the weak-coupling limit when the gap is small compared to the
electron band-width.~\cite{grage, controzzi} 
At finite excitation energies
exact analytical results are available in the strong-coupling 
regime~\cite{penc-spectral, pencdensity} or
when additional symmetries are present.~\cite{arikawa}

There are several numerical approaches for
calculating momentum- and frequency-dependent correlation functions in
lattice models of correlated electrons. These
include Exact Diagonalization,~\cite{bannister} 
Cluster Perturbation Theory (CPT),~\cite{senechal}
Variational CPT,~\cite{potthoff} 
and Quantum Monte Carlo algorithms.~\cite{preuss,lavalle} 
Here we employ the Dynamical Density-Matrix Renormalization Group
(DDMRG) method.~\cite{ddmrg} 
It is a zero-temperature method which has been successfully 
applied to the study of dynamical properties in
metallic and insulating phases of one-dimensional
extended Hubbard models.~\cite{rixs,hubbard-spectral,tioclpaper,diss,matsueda}

In this paper we consider the dynamical properties of
the half-filled extended Hubbard model with nearest-neighbor interaction. 
It is believed to be a minimal model for
quasi-one-dimensional Mott-insulators, such as the cuprate compounds
SrCuO$_2$ and Sr$_2$CuO$_3$. 
In a previous work~\cite{rixs}
we have shown that this model describes the optical conductivity 
and the dispersive structures observed in the resonant inelastic x-ray
scattering
(RIXS) experiments of SrCuO$_2$. Here we show that the model also
captures essential
results obtained in inelastic neutron scattering,~\cite{zaliznyak}
electron energy loss spectroscopy (EELS),~\cite{eels}
and ARPES~\cite{kim,koi06}
experiments. To this end we present numerical results for the
energy- and momentum-resolved spin and charge structure factors and the
one-particle spectral function of the extended Hubbard model. 
We also discuss in detail the influence of the model parameters on the line
shapes
of structure factors and spectral functions.

\section{Model and Methods}

The one-dimensional extended Hubbard model is defined by the
Hamiltonian 
\begin{eqnarray}
\hat{H} 
&=& -t \sum_{l,\sigma} \left( \hat{c}^{\dagger}_{l,\sigma} 
\hat{c}^{\phantom{\dagger}}_{l+1,\sigma} 
+ \hat{c}^{\dagger}_{l+1,\sigma} \hat{c}^{\phantom{\dagger}}_{l,\sigma} \right)
\nonumber \\
&&+ U \sum_{l} \left(\hat{n}_{l,\uparrow}-\frac{1}{2}\right)
\left(\hat{n}_{l,\downarrow}-\frac{1}{2}\right)
\label{Hamiltonian} \\
&&+ V \sum_{l} \left( \hat{n}^{\phantom{\dagger}}_{l} -1 \right)
\left( \hat{n}^{\phantom{\dagger}}_{l+1} -1 \right)  . \nonumber 
\end{eqnarray}
Here $\hat{c}^{\dagger}_{l,\sigma}$, $\hat{c}^{\phantom{\dagger}}_{l,\sigma}$ 
are fermionic creation and 
annihilation operators for a particle with spin 
$\sigma = \uparrow,\downarrow$ at site $l=1,\dots,L$, 
$\hat{n}^{\phantom{\dagger}}_{l,\sigma}= 
\hat{c}^{\dagger}_{l,\sigma}\hat{c}^{\phantom{\dagger}}_{l,\sigma}$, 
and $\hat{n}^{\phantom{\dagger}}_l=
\hat{n}^{\phantom{\dagger}}_{l,\uparrow}+\hat{n}^{\phantom{\dagger}}_{l,
\downarrow}$.
In the following we consider chains with open boundary conditions
and an even number of lattice sites. 
The total number
of electrons is equal to the number of lattice sites (half-filled band).

For $V/t=0$ the model reduces to the simple Hubbard Hamiltonian. Without
magnetic field the physical excitations of the insulating Hubbard model are
combinations of  collective modes, spinons and 
(anti-)holons, for any $U>0$.~\cite{theBook}
The spinons are gapless and charge neutral
excitations that carry spin $1/2$. Anti-holons and holons
are gapped modes with charge $\pm e$ and no spin. 
When the nearest-neighbor interaction $V$ is turned on and $U > 0$,
the system remains
a Mott-insulator until $V \approx U/2$. For larger $V$ the system becomes a
charge-density wave (CDW) insulator.~\cite{eric-ehm} In this work we only
consider the Mott-insulating regime $U > 2V \geq 0$.
The parameters $U=7.8t, V=1.3t$ and $t=0.435eV$ were shown in 
Ref.~\onlinecite{rixs} to describe both the
optical conductivity and the dispersive structures observed in the RIXS spectrum
for SrCuO$_2$.
In addition these parameters yield an effective spin-exchange
coupling $J\approx 0.24{\rm eV}$ which is in good agreement with recent
high-precision neutron scattering data.~\cite{zaliznyak}

The dynamical spin structure factors $S(q,\omega)$
is the imaginary part of the spin-spin 
correlation function 
\begin{equation}
S(q,\omega) = \frac{1}{\pi} \text{Im}
\langle \psi_0| \hat{S}_{q}^{z\dagger} 
\frac{1}{\hat{H}+\omega-E_0-i\eta} \hat{S}_{q}^{z}
| \psi_0 \rangle  ,\label{eq:spinstruct}
\end{equation}
where $|\psi_0\rangle$ and $E_0$ are the ground state wave function and energy
and $\eta \rightarrow 0^+$.
The operator 
$\hat{S}_{q}^z$ is the Fourier transform of the local z-component of the
spin operator 
$\hat{S}_l^z=\hat{n}_{l,\uparrow}-\hat{n}_{l,\downarrow}$. 
We set $\hbar = 1$ everywhere, so that an excitation momentum and energy are 
equal to the wavevector $q$ and the frequency $\omega$, respectively.
When we replace
$\hat{S}_{q}^z$ by the Fourier transform $\hat{n}_q$ of the local particle 
density
$n_l=\hat{n}_{l,\uparrow}+\hat{n}_{l,\downarrow}$ we obtain the dynamical charge
structure factor 
\begin{equation}
N(q,\omega) = \frac{1}{\pi} \text{Im}
\langle \psi_0| \hat{n}^{\dagger}_{q}
\frac{1}{\hat{H}+\omega-E_0-i\eta} \hat{n}_{q}
| \psi_0 \rangle  .\label{eq:chargestruct}
\end{equation}
Similarly, the one-particle spectral function is defined through
\begin{equation}
A(q,\omega) = \frac{1}{\pi} \text{Im}
\langle \psi_0| \hat{c}_{q, \sigma}^{\dag} 
\frac{1}{\hat{H}+\omega-E_0-i\eta} \hat{c}_{q,\sigma}^{\phantom{\dagger}}
| \psi_0 \rangle  ,\label{eq:spectral}
\end{equation}
where $\hat{c}_{q, \sigma}^{\phantom{\dagger}}$ is the Fourier transform of the
annihilation operator $\hat{c}_{l, \sigma}^{\phantom{\dagger}}$.

These dynamical correlation functions can be calculated with the DDMRG
method for a finite system size $L$ and a finite broadening $\eta$.~\cite{ddmrg}
Therefore, all DDMRG spectra presented here are a convolution of the true
dynamical correlation function with
a Lorentzian distribution of width $\eta$. 
The spectral properties in the thermodynamic limit can be determined using a
finite-size scaling analysis with an appropriate 
size-dependent broadening $\eta(L)\propto 1/L$. 
In general, momentum dependent operators 
$\hat{a}_{q}=\hat{S}_{q}^z, \hat{n}_{q},$ or 
$\hat{c}_{q, \sigma}^{\phantom{\dagger}} $ are defined from
local operators $\hat{a}_{l} = \hat{S}_{l}^z, \hat{n}_{l},$ or 
$\hat{c}_{l, \sigma}^{\phantom{\dagger}}$
using the one-electron eigenstates
of the non-interacting system with 
periodic boundary conditions, i.e.
\begin{equation}
\hat{a}_{q} = \frac{1}{\sqrt{L}} \sum_{l} e^{-iql} \hat{a}_{l}
\end{equation}
with momentum $q=2\pi Z/L$ 
(we set the lattice constant equal to 1) 
and integers $-L/2< Z\leq L/2$.
Since the DMRG algorithm performs best for open boundary conditions,
we extend the
definitions of $S(q,\omega)$, $N(q,\omega)$, and $A(q,\omega)$ and write
\begin{equation}
\hat{a}_{q} = \sqrt{\frac{2}{L+1}} \sum_{l} \sin(kl) \hat{a}_{l}
\end{equation}
with (quasi-)momenta $q=\pi Z/(L+1)$ for integers $ 1 \leq Z \leq L$.
The expansion coefficients are the eigenstates of a free particle in
a box.
Both definitions of $\hat{a}_{q}$ should become equivalent in the
thermodynamic limit $L\rightarrow \infty$.
Comparisons with Bethe Ansatz results for the Hubbard model
have confirmed that this procedure gives accurate 
results for $q$-dependent quantities.~\cite{hubbard-spectral,diss,tioclpaper}
In our DDMRG calculations we have used up to $400$ density matrix eigenstates.
For all results presented here truncation errors are negligible compared to the
finite resolution in momentum [$\Delta q = \pi/(L+1)$] and in energy 
($\Delta \omega \sim \eta \propto 1/L$) imposed by finite system sizes.

\section{Dynamical Spin Structure Factor}

In this section we present our results
for the dynamical spin structure factor
$S(q,\omega)$.
We show that the spin dynamics of the extended Hubbard model
agrees with a recent high-precision inelastic
neutron scattering experiment
in SrCuO$_2$.~\cite{zaliznyak} In addition, we discuss the influence of the
model parameters on the spin structure factor of the
Hamiltonian~(\ref{Hamiltonian}).

\begin{figure}
\includegraphics[width=5cm]{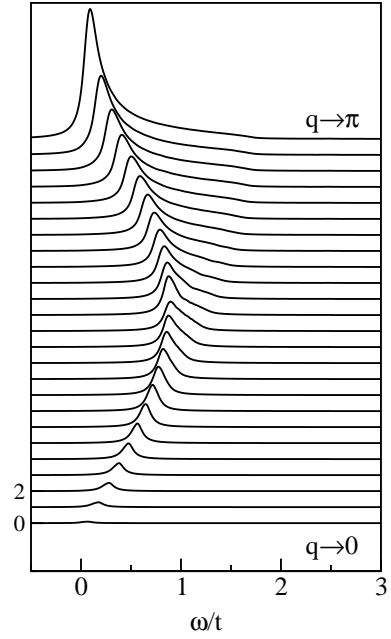}
\caption{
Line shapes of the spin structure factor
$S(q,\omega)$ for $0<q<\pi$
calculated with a broadening $\eta = 0.05t$
for $U=7.8t$ and $V=1.3t$.
}
\label{fig:lineshapes}
\end{figure}

Figure~\ref{fig:lineshapes} shows the line shapes of the dynamical spin 
structure
factor in the extended Hubbard model for $U=7.8t$ and $V=1.3t$.  
The system size is $L=100$ and we have used
a broadening $\eta = 5t/L$.
The spectrum is dominated by the
spectral weight at the lower onset which disperses from zero at $q=0,\pi$ to a
maximum at $\pi/2$. Above the low energy peak there is a continuum of spectral
weight which is bounded by an upper onset. 

These line shapes are strongly
reminiscent of the spin structure factor of the spin-1/2 Heisenberg chain, which
in good approximation is described by the M\"{u}ller-Ansatz~\cite{mueller}. The
M\"{u}ller-Ansatz structure factor is given by
\begin{equation}
S_{\rm MA}(q,\omega) = A \frac{\theta(\omega-\omega_{\rm L}(q)
)\theta(\omega_{\rm U}(q) -\omega )}
{\sqrt{\omega^2 - \omega^2_{\rm L}(q)}}\label{eq:mueller-ansatz}
\end{equation}
where $A$ is a prefactor and $\omega_{\rm U,L}(q)$ 
are the des Cloiseaux-Pearson (dCP) dispersion relations.
They describe the compact support of the two-spinon continuum of the spin-1/2
Heisenberg model
\begin{subequations}
\begin{eqnarray}
\omega_{\rm L}(q) &=& \frac{\pi J}{2}|\sin(q)|\, \label{eq:lower-dCP}\,,\\
\omega_{\rm U}(q) &=& \pi J|\sin(q/2)|\;. \label{eq:upper-dCP}
\end{eqnarray}
\end{subequations}
where $J$ is the nearest-neighbor spin-exchange coupling.
The M\"{u}ller-Ansatz structure factor is non-zero only within the bounds of
$\omega_{\rm U,L}(q)$ and there is a square-root divergence at the low-energy
onset.

\begin{figure}
\includegraphics[width=6.5cm]{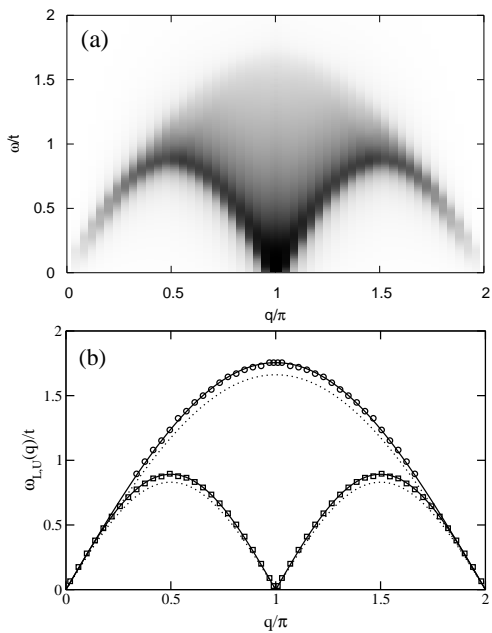}
\caption{
(a) Density plot of the data in Fig.~\ref{fig:lineshapes}.
(b) Upper (circles) and lower (squares) onset of the two-spinon continuum
extracted from the DDMRG data. Solid lines are fits to the dCP dispersion 
relations.
The dotted line is the dCP dispersion using the exchange coupling
$J_{\rm Exp}=0.226$ eV from neutron scattering experiments in
SrCuO$_2$.~\cite{zaliznyak}
}
\label{fig:dens-and-onset}
\end{figure}

The similarity of $S_{\rm MA}(q,\omega)$ with $S(q,\omega)$ in the extended
Hubbard model can be seen more clearly in a density plot.
Figure~\ref{fig:dens-and-onset}(a) shows a density plot of the
spectral weight distribution. Most spectral weight is located at the lower
onset of the two-spinon continuum
as in the Heisenberg model. Figure~\ref{fig:dens-and-onset}(b)
shows the lower and upper boundaries of the two-spinon continuum
extracted from the DDMRG data. 
The upper boundary was obtained by analyzing the second
derivative ${\mathrm d}^2S(q,\omega)/{\mathrm d}\omega^2$ and the lower 
onset was identified with the position of the low-energy peak. 
These boundaries are very similar to the 
dCP--dispersion relations (\ref{eq:lower-dCP}) and (\ref{eq:upper-dCP}) of
the Heisenberg model. 
Fitting our numerical data to these relations we obtain 
effective exchange couplings $J=0.242{\rm eV}$ (lower boundary)
and $J=0.248{\rm eV}$ (upper boundary), respectively (using $t=0.435$ eV). 
These values are in good agreement with the value $J=0.24{\rm eV}$
reported in our previous work~\cite{rixs} and with the 
coupling $J_{\rm Exp}= 0.226{\rm eV}$ determined from inelastic
neutron scattering data for SrCuO$_2$.~\cite{zaliznyak}
As the M\"{u}ller-Ansatz (\ref{eq:mueller-ansatz}) 
with the exchange coupling $J_{\rm Exp}= 0.226{\rm eV}$ 
describes the neutron scattering spectrum of SrCuO$_2$
very well,~\cite{zaliznyak} we conclude that the experimental 
spectrum is also well described by the spin structure factor 
of the extended Hubbard model for $U=7.8t, V=1.3t$ 
and $t=0.435$ eV. 

\begin{figure}
\includegraphics[width=8cm]{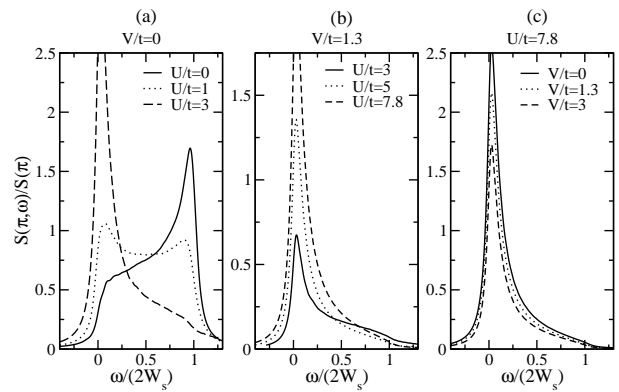}
\caption{
Dependence of the spin structure factor at momentum $q\approx\pi$ on 
$U$ and $V$.  
The broadening is $\eta=W_s/20$ in the left panel and $\eta=W_s/10$ in the
other two panels.
}
\label{fig:peak-at-pi}
\end{figure}

A recent study~\cite{grage} of the spin structure factor in the
Hubbard model has shown that there are significant itinerancy effects
[a transfer of spectral weight in $S(q,\omega)$ due to the coupling of the
spin excitations to charge fluctuations]
at low and intermediate values of the Hubbard interaction $U$.
In this regime the spin structure factor of the Hubbard model 
differs from the spectrum $S_{\rm MA}(q,\omega)$ of the Heisenberg
model (i.e., the strong-coupling limit $U \gg t$ of the Hubbard model), 
where charge fluctuations are
suppressed and electrons are completely localized. 
To illustrate this effect we show DDMRG
spectra obtained for various parameters $U$ and $V$ on 60-site lattices 
in Fig.~\ref{fig:peak-at-pi}.
We have normalized $S(q\approx\pi,\omega)$
by the total spectral weight $S(\pi)=\int S(\pi,\omega){\rm d}\omega$,
which we obtain by calculating the ground state 
expectation value $S(\pi)=\langle \psi_0 \rvert S_{q=\pi}^z S_{q=\pi}^z \vert
\psi_0 \rangle$.
In these figures the energies are given in units of the two-spinon band width
$2 W_s$, which is equal to $\pi J$ in the Heisenberg model.
For $V/t=0$ the value of $W_{\rm s}$ is known exactly from the thermodynamic 
Bethe-Ansatz solution.~\cite{theBook}
When $V/t\neq 0$ we determine $W_{\rm s}$ from the upper onset of the 
dynamical spin structure factor.

Figure~\ref{fig:peak-at-pi}(a) shows $S(\pi,\omega)$ 
calculated with a broadening $\eta = W_s/20$
for $V/t=0$ and three values of $U/t=0,1,3$.
At $U/t=0$ we recover
the case of free electrons and accordingly most spectral weight is found
in the peak at the upper onset ($\omega = 2W_s$) of $S(q,\omega)$.
When $U/t$ becomes larger
the spectral weight is quickly redistributed towards the lower onset
at $\omega = 0$. Already
at $U/t=3$ the upper peak is no longer visible and most spectral weight
is found at the lower onset. 
This supports the analysis of Ref.~\onlinecite{grage} but shows
that itinerancy effects become rapidly weaker for increasing $U$.
We now set $V/t=1.3$ and vary $U/t$. The resulting curves
$S(\pi,\omega)$ are shown in Figure~\ref{fig:peak-at-pi}(b). 
At $U/t=3$ the largest part of the
spectral weight is at the low-energy onset. With increasing $U/t$
the low-energy peak becomes more and more pronounced until only little
spectral weight remains at the upper onset.
For larger interaction strengths the change of the next-nearest neighbor
interaction has a weaker influence on the spectral distribution. 
Figure~\ref{fig:peak-at-pi}(c) shows $S(\pi,\omega)$ for constant $U/t=7.8$ and
varying $V/t$. While the weight of the peak at $\omega=0$ is 
clearly diminished with growing $V$, the effect appears to be less
pronounced than in the case of smaller $U$.
In summary, our DDMRG results for the extended Hubbard model confirms 
that itinerancy effects are important in the dynamical spin structure factor 
of Mott insulators.~\cite{grage} 
We note, however, that despite the coupling between spin and charge sectors
the upper and lower onsets of the two-spinon continuum
are very well described 
by the dCP relations~(\ref{eq:upper-dCP}) and (\ref{eq:lower-dCP}) of
the Heisenberg model from strong to intermediate interaction
as illustrated by the results in Fig.~\ref{fig:dens-and-onset}.

\begin{figure}
\includegraphics[width=8cm]{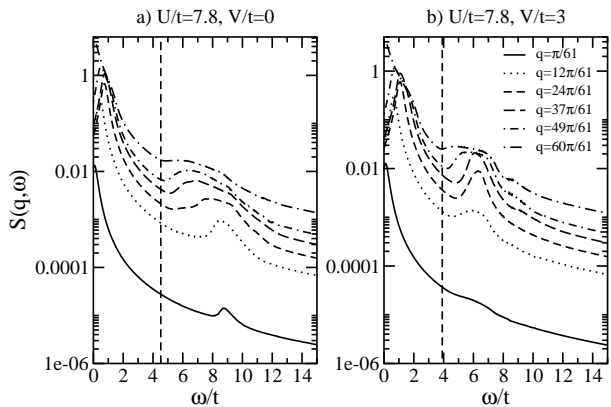}
\caption{
Spin structure factor $S(q,\omega)$ 
at high energies for
(a) $U/t=7.8$, $V/t=0$ and (b) $U/t=7.8$, $V/t=3$. The broadening is
$\eta =0.2t$.  Vertical lines indicate the one-particle gaps $\Delta_c$.
}
\label{fig:hi-energy-spin-corr}
\end{figure}

Our numerically results reveals another effect of charge fluctuations.
The total spectral weight of the spin structure factor is not exhausted by 
low-energy excitations $\omega(q) \leq \omega_{\text U}(q)$.   
There are dispersive features in that spectrum at 
energies above the one-particle gap, $\Delta_c=E_0(1)-E_0(1)+2E_0(0)$,
where $E_0(x)$ denotes the ground state energy of the 
Hamiltonian~(\ref{Hamiltonian})
with $x$ electrons added to or removed from the half-filled system.
Figure~\ref{fig:hi-energy-spin-corr} shows the high energy spectrum of 
$S(q,\omega)$ for $U/t=7.8$, $V/t=0$ and $U/t=7.8$, $V/t=3$
on a logarithmic scale ($L=60$ in both cases). 
For energies
higher than the one-particle gap we can see that there are dispersive structures
up to an energy of roughly twice the one-particle gap. The presence of
these spectral features further shows that the coupling to the 
charge sector affects the spin dynamics of the system.

\section{Dynamical Charge Structure Factor}

In this section we present numerical results for the dynamical
charge structure factor $N(q,\omega)$ of the extended Hubbard model. 
This correlation function
is probed by electron-electron energy loss spectroscopy (EELS). It
is also believed to describe the dispersion of peaks and onsets 
in RIXS experiments~\cite{rixs}.

\begin{figure}
\includegraphics[width=8cm]{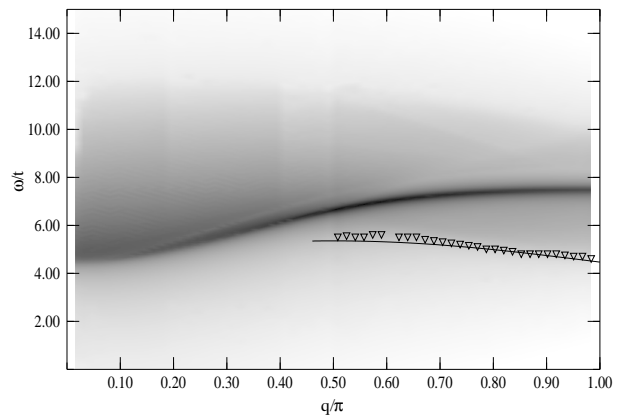}
\caption{
Density plot of the renormalized 
charge structure factor $N(q,\omega)/q^2$
for $U/t=7.8$, $V/t=1.3$. For $q \agt \pi/2$ a solid line follows
the spinon dispersion (shifted by the value of the 
charge gap) and triangles indicate
the position of the spectrum onset in the numerical data.
}
\label{fig:nqw-dens}
\end{figure}

In Fig.~\ref{fig:nqw-dens} we show a density plot of $N(q,\omega)/q^2$ for the
parameters $U/t=7.8$, $V/t=1.3$ relevant for SrCuO$_2$. The present 
DDMRG data are significantly better resolved than in our  previous 
work~\cite{rixs}.
Since $\mathrm{lim}_{q\to0}\int \mathrm{d}\omega \,N(q,\omega)\propto q^2$
we have plotted $N(q,\omega)/q^2$ on a logarithmic scale in 
Fig.~\ref{fig:nqw-dens}.
The momentum resolution is $\Delta q/\pi=1/61$ and the energy resolution 
is $\eta/t=0.1$ 
for a system of $L=60$ sites. 
The most  prominent feature is a resonance  that disperses downward
monotonically starting at the Brillouin zone boundary $q = \pi$.
In the strong-coupling limit~\cite{pencdensity} the exciton 
lies below the continuum for all momenta. 
Here, in contrast, there is also a substantial
spectral weight below the resonance, which clearly lies
above the low-energy onset of the continuum shown 
by triangles for $q \geq \pi/2$ in Fig.~\ref{fig:nqw-dens}. 
As already found in Ref.~\onlinecite{rixs} this low-energy onset of the continuum 
follows the spinon dispersion (shifted by the value of the charge gap),
which is indicated by a line for $q \agt \pi/2$ in Fig.~\ref{fig:nqw-dens}.

\begin{figure}
\includegraphics[width=8cm]{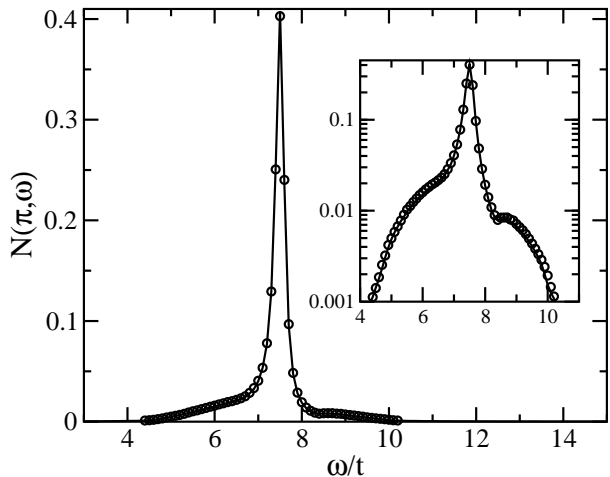}
\caption{Charge structure factor at $q=\pi$ for $U/t=7.8$ and $V/t=1.3$ 
calculated
with a broadening $\eta/t=0.1$. 
System lengths are $L=60$ (full line) and $L=120$ (open circles).
The inset shows the same data on a logarithmic scale.
\label{fig:peakatpi}
}
\end{figure}

The presence of spectral weight below the resonance is particularily blatant
at the Brillouin zone boundary. This
can be seen clearly 
in Fig.~\ref{fig:peakatpi}, where we show
spectra  $N(q\approx\pi,\omega)$ calculated for 60 and 120 lattice sites
and the
same broadening $\eta/t=0.1$. Both spectra are almost indistinguishable which
excludes the possibility that the observed spectral weight is a finite-size
effect. In the inset
of Fig.~\ref{fig:peakatpi} the same data are shown on a logarithmic scale.
We note that more spectral weight lies below the resonance than above it.
Just  above the resonance there is a small
dip in $N(q\approx\pi,\omega)$
reminiscent of an asymmetric Fano line shapes.~\cite{fano} 

In our previous work~\cite{rixs} we reported that the extended Hubbard model
with $U/t=7.8$, $V/t=1.3$ and $t= 0.435$ eV gives an accurate description
of the low-energy optical conductivity 
$\sigma(\omega)=\omega\mathrm{lim}_{q\to0} N(q,\omega)/q^2$
of SrCuO$_2$. Furthermore, we showed that the low-enery onset and peak 
dispersions in $N(q,\omega)$ are in quantitative agreement with 
similar dispersive structures observed in a RIXS experiment.
We can also compare our DDMRG results for $N(q,\omega)$ in the extended Hubbard 
model to the EELS spectrum of Sr$_2$CuO$_3$.~\cite{eels}
As the model parameters for this material
are close to those for SrCuO$_2$ (see below), we expect
to observe the same qualitative features in both cases. 
Actually, we find that our results for the charge structure factor
are in qualitative agreement with the 
EELS spectrum of Sr$_2$CuO$_3$.
In particular, the low-energy tail of $N(q,\omega)$ explains 
the signal observed experimentally below the resonance close to the
zone boundary,~\cite{moskvin} a feature which could not be explained
by the strong-coupling theory.

\begin{figure}
\includegraphics[width=8cm]{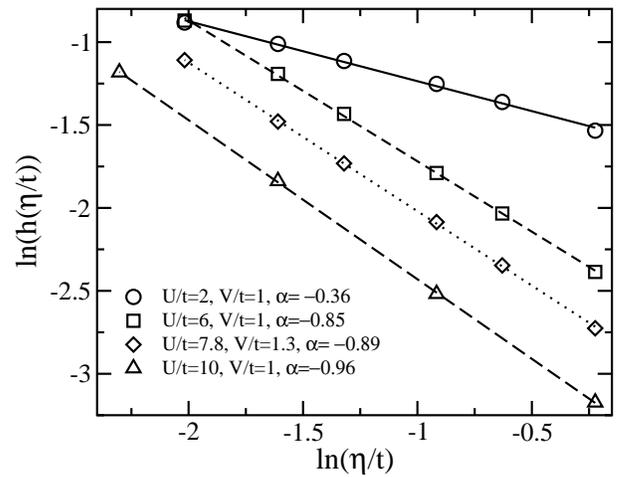}
\caption{
Scaling analysis of the logarithm of the
peak height $h$ in $N(q\approx\pi,\omega)$ as 
a function of the logarithm of the
size-dependent broadening $\eta(L)=12t/L$ for various
interaction strengths. Lines are linear fits.}
\label{fig:scaling}
\end{figure}

To understand the nature of the resonance observed in  $N(q\approx\pi,\omega)$
we have performed a finite-size analysis of the peak height $h(\eta(L))$
for different interaction strengths. 
The broadening is size-dependent such that $\eta L=12t$.~\cite{ddmrg}
In strong-coupling theory~\cite{pencdensity} the dispersive resonance is a
$\delta$-peak indicating the presence of an exciton. 
As discussed in Ref.~\onlinecite{ddmrg} the peak height should diverge as 
$h(\eta)\propto \eta^{-1}$ in that case. 
Figure~\ref{fig:scaling} shows a plot of $h(\eta)$ on a double
logarithnic scale for various couplings. 
Clearly,  $h(\eta)$ diverges as power law $\eta^\alpha$ for 
$\eta \rightarrow 0$. 
We extract the exponents $\alpha$ using a linear fit of the logarithms
of $h(\eta)$ and $\eta$.  
The error in the value of the exponent $\alpha$ is
0.01 or smaller.
For $U/t \leq 10$ we clearly find
that $0 > \alpha > -1$, which indicates a power-law
divergence in the spectrum $N(q,\omega)$ but no $\delta$-peak.~\cite{ddmrg}
In fact, the prediction of the strong-coupling expansion,
$\alpha \approx -1$, 
is only reached (within the numerical accuracy) for extremely large
$U/t \geq 100$ (not shown in Fig.~\ref{fig:scaling}).
This is a first indication that strong-coupling theory becomes
accurate only for unphysically large on-site Coulomb interaction. 

Since the onset of $N(q,\omega)$ follows the spinon dispersion,
it is clear that the excited states contributing to the charge structure factor
consist in charge excitations hybridized with spin excitations. 
This is the reciprocal effect of the itinerancy effect in the
spin structure factor.
The Fano-type line shape of the resonance suggests that it
originates from a discrete charge excitation coupled to a continuum of 
spin excitations. In the strong-coupling limit this discrete
charge excitation is an exciton (a neutral bound state of a holon and an 
anti-holon). As the spin excitation band width goes to zero
in that limit, only elastic scattering is possible and 
thus the exciton has an infinite lifetime and appears
as a $\delta$-peak in the spectrum $N(q,\omega)$.
For a finite interaction strength, however, spin excitations
have a finite band width. Thus neutral bound states have
a decay channel into independent charge excitations
due to the (inelastic) scattering by spin excitations 
and therefore their lifetime is finite. Accordingly, 
the resonance is no longer a $\delta$-peak but a power-law divergence.
In particular, for the couplings $U$ and $V$ relevant for real materials,
such as the Mott-insulating chain cuprates SrCuO$_2$ and Sr$_2$CuO$_3$,
the resonance is not an exciton but corresponds to neutral
bound states with a finite lifetime.

\begin{figure}
\includegraphics[width=8cm]{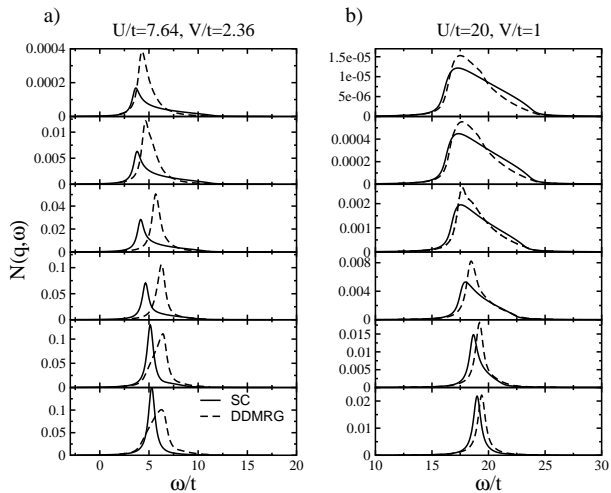}
\caption{
Comparison between the strong-coupling~\cite{pencdensity} (solid lines)
and DDMRG (dashed lines)
results for the charge structure factor for two sets of model
parameters. 
A broadening $\eta/t=0.4$ is used for both approaches.
The momentum varies from $q =  0.03\pi$  (top) to
$q =  0.97\pi$  (bottom).
} 
\label{fig:sc-vs-ddmrg}
\end{figure}

As a further check of the strong-coupling theory
we directly compare its predictions with our DDMRG results
in Fig.~\ref{fig:sc-vs-ddmrg}.
The DDMRG data were obtained for a $30$-site lattice and a quite large
broadening $\eta/t=0.4$, which smears out fine details.
We have convolved the  strong-coupling structure factor with a Lorentzian
of the same width $\eta$ to enable a direct comparison.

In Fig.~\ref{fig:sc-vs-ddmrg}(a) we show $N(q,\omega)$ for 
$U/t=7.64$, $V/t=2.36$.
These parameters were chosen in Ref.~\onlinecite{eels}
to explain EELS data of Sr$_2$CuO$_3$
using the strong-coupling approach.
However, the comparison in Fig.~\ref{fig:sc-vs-ddmrg}(a) reveals
large deviations between DDMRG and strong-coupling results for these
parameters.
Therefore, our results demonstrate that a strong-coupling approach 
is not accurate in this parameter regime and the parameters 
$U/t=7.64$, $V/t=2.36$ determined using this approach are not reliable.
For comparison, a fit of the experimental
optical conductivity and exchange coupling to
DDMRG results for the extended Hubbard model (as done 
in Ref.~\onlinecite{rixs} for SrCuO$_2$)
yields the parameters $U/t=7.4$, $V/t=1.8$, and
$t=0.42$ eV for Sr$_2$CuO$_3$.
We note that the apparently small difference between strong-coupling and DDMRG
parameters has actually radical effects, as the optical absorption
spectrum contains an exciton below the Mott gap
for $V > 2t$ but not for $V \leq 2t$.
(Unfortunately, due to the finite experimental resolution it is not clear
whether there is such an exciton in the linear optical absorption of
Sr$_2$CuO$_3$.) 

As expected the agreement between strong-coupling theory and DDMRG becomes
better for stronger interactions. This can be seen in 
Fig.~\ref{fig:sc-vs-ddmrg}(b), where we compare $N(q,\omega)$ for $U/t=20$ and
$V/t=1$. Nevertheless, in spite of the interaction equaling five times the 
bare band width $4t$, we still observe significant
deviations from the strong-coupling predictions. 
Similar discrepancies were found previously at coupling as large as
$U=40t$ in a study of the optical 
conductivity,~\cite{eric-mott} which is related to $N(q,\omega)$ through 
$\sigma(\omega)/\omega=\mathrm{lim}_{q\to0} N(q,\omega)/q^2$.

\begin{figure}
\includegraphics[width=8cm]{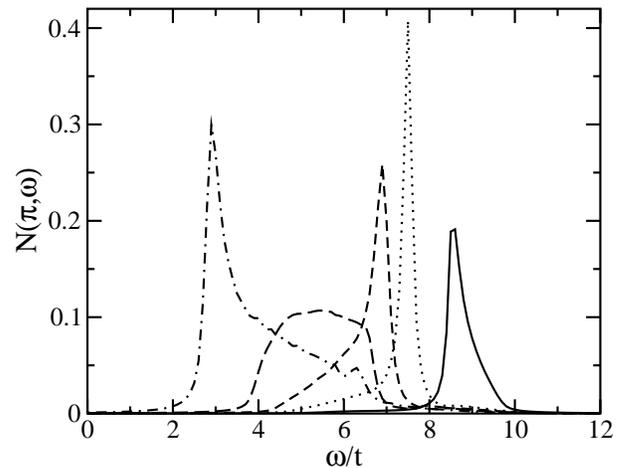}
\caption{
Charge structure factor $N(q\approx\pi,\omega)$ for $U/t=7.8$ and 
$V/t=0$ (full line), $V/t=1.3$ (dotted line), $V/t=2$ (short dashed line),
$V/t=2.75$ (long dashed line), and $V/t=3.5$ (dashed dotted line).
A broadening $\eta = 0.1t$ is used.
}
\label{fig:evolution}
\end{figure}

It has been shown in Ref.~\onlinecite{eric-optics-ehm} that there are four
different
types of optical excitations that contribute to $\sigma(\omega)$
depending on the strength of the next-neighbor repulsion $V$. We now study 
the evolution of $N(q=\pi,\omega)$ with $V$ and discuss how the nature
of these excitations influences the dynamical charge structure factor.
Figure~\ref{fig:evolution} shows $N(\pi,\omega)$ for $U/t=7.8$ and 
several values of $V/t$. As expected, we find a broad peak when there is
no next-neighbor interaction since the charge excitations do not
form long-lived bound states. 
For $V/t=1.3$ we find a sharp peak with a small but finite intrinsic
width. When $V/t=2$ the resonance is weaker and spectral weight leaks
to lower energies which leads to a pronounced asymmetry of the peak. This
trend continues until the peak is no longer visible in the continuum.
The spectral weight is now distributed in a broad band as seen
for $V/t=2.75$ in Fig.~\ref{fig:evolution}. In the optical conductivity
CDW droplets are known to give rise to such a band.~\cite{eric-optics-ehm}
Therefore we suggest that the band observed at $q=\pi$ is also related to 
CDW droplets.
When $V$ is further increased
a peak appears at low energies which contains most spectral weight.
Approaching the CDW phase boundary ($V_c/t\approx3.5$ for $U/t=7.8$) 
the total spectral weight in that peak 
$N(\pi)=\int\,{\rm d}\omega N(\pi,\omega)$ (i.e, the
static charge structure factor at $q=\pi$)
becomes extremely large, which signals the occurrence of the long-range CDW
order for $V > V_c$.

\section{One-Particle Spectral Function}

\begin{figure}
\includegraphics[width=8cm]{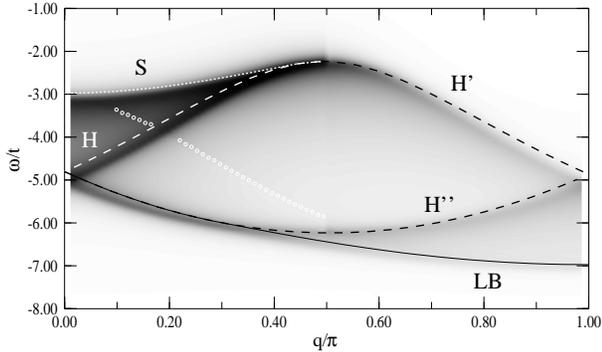}
\caption{
Density plot of the one-particle spectral function $A(q,\omega)$
calculated with DDMRG  for $U/t=7.8$ and $V/t=1.3$. 
The dispersions
of elementary excitations in the one-dimensional
Hubbard model with $U/t=7.74$ are also shown: Spinon branch (S), holon branches 
(H,H',H'') and lower boundary of the holon-spinon continuum (LB).
Open white circles mark the position of a step-like feature
in the DDMRG spectrum.  
}
\label{fig:greensfn}
\end{figure}

We now consider the one-particle spectral function $A(q,\omega)$ of the
Hamiltonian~(\ref{Hamiltonian}). This dynamical correlation function
corresponds to the spectrum measured in ARPES experiments. 
A density plot of the spectral function is shown in Fig.~\ref{fig:greensfn}
for the parameters appropriate for SrCuO$_2$, $U/t=7.8$ and $V/t=1.3$. 
These data have been calculated on a 90-site lattice using a broadening
$\eta=0.1t$, which yields a momentum resolution $\Delta q=\pi/91$
and an energy resolution of the order of $\eta$.
With this resolution spin-charge separation is clearly seen
in Fig.~\ref{fig:greensfn}. The spinon branch is denoted by (S)
and has a width $W_s = \pi J/2 \approx 0.9t \approx 0.4$ eV.
The holon branch is labeled (H) and has a width of about 
$2.7t \approx 1.2$ eV (from $q=0$ to $q=\pi/2$).
This spin-charge separation has been observed experimentally
in the ARPES spectrum of SrCuO$_2$.~\cite{kim,koi06} 
The spinon and holon band widths calculated here for $U/t=7.8$ and $V/t=1.3$
are compatible with the experimental data: 
The latest measurements~\cite{koi06} yield $0.49 \pm 0.13$ eV and
$1.42 \pm 0.08$ eV, respectively.
[Note that the effective hopping term $t$ defined in Ref.~\onlinecite{koi06}
is not equivalent to the bare hopping term $t$ of our 
model~(\ref{Hamiltonian}).]
A further verification of the theoretical description of SrCuO$_2$
by the extended Hubbard model would be a direct comparison
of the experimental spectral weight distribution with our DDMRG data.
Unfortunately, the limited resolution of ARPES data
and a strong background signal prevent a meaningful comparison.

In addition to the holon and spinon branches one sees
several dispersive structures in the spectral function of the extended
Hubbard model, which are quite similar to those found in the Hubbard
model ($V=0$).~\cite{penc-spectral,senechal,matsueda} 
To identify these structures we have calculated the exact dispersions 
of various excitations in a periodic 90-site Hubbard chain using
the Bethe Ansatz. 
The parameter $U/t=7.74$ has been chosen to reproduce
the one-particle gap of the extended Hubbard model with $U/t=7.8$ and
$V/t=1.3$.
The dispersions of the Hubbard model corresponding to the most
important features in
the spectral function of the extended Hubbard model
are shown as lines in Fig.~\ref{fig:greensfn}. There are 
the spinon branch (S), the holon branch (H), a secondary holon branch (H'),
the continuation of the holon branch (H''), 
and the lower boundary (LB) of the continuum of excitations
consisting of a single spinon and a single holon .
In the entire Brillouin zone these dispersions of the Hubbard model
follow closely the dispersive structures of the extended Hubbard model. 
The deviations can be mainly attributed to a change
of the spinon and holon band widths.
Thus we conclude that in the Mott-insulating regime
the spectral functions of the extended Hubbard model are qualitatively similar
to those of the Hubbard model.

There is a very weak dispersive structure in $A(q,\omega)$ of the extended
Hubbard model which is not visible in the density plot and is
indicated by open circles in Fig.~\ref{fig:greensfn}. 
This feature is not present in the
spectral function of the Hubbard model with $U/t=7.74$.
In the extended Hubbard model it consists of a shoulder in the spectrum and
can be localized by calculating the first derivative of $A(q,\omega)$. 
Its dispersion closely resembles the holon dispersion shifted by an offset
equal to the spinon band width.
However, we do not find this shoulder if we use periodic boundary conditions.
Therefore, this feature is a boundary effect due to the abrupt cutting-off
of the nearest-neighbor interaction terms at both chain ends.

We note that the spectral function $A(q,\omega)$ of the extended Hubbard
model has been investigated previously using a variational CPT 
approach~\cite{potthoff} for a set of parameters ($U/t=8, V/t=2$) 
close the one used in Fig.~\ref{fig:greensfn}.
Qualitatively, both spectral functions are very similar and resemble the 
spectral function of the Hubbard model.~\cite{penc-spectral,senechal,matsueda} 
However, our DDMRG spectra are free of the 
striped structures which are present in the variational CPT spectral functions.
Therefore, we conclude that these striped structures are artifact of the
variational CPT approach.

\begin{figure}
\includegraphics[width=8cm]{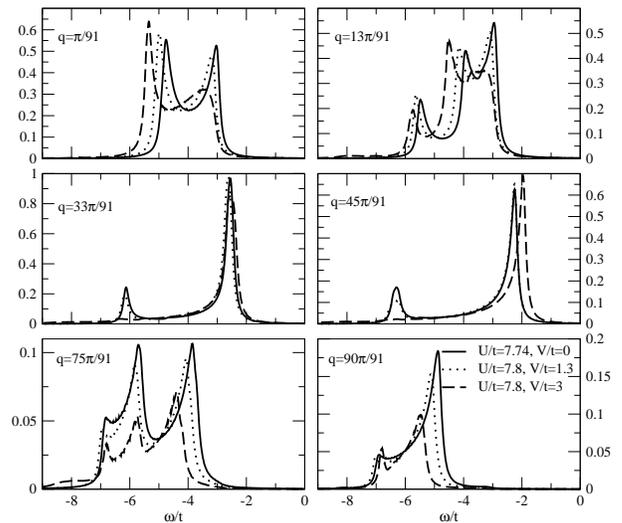}
\caption{
One-particle spectral function $A(q,\omega)$ for various 
momenta $q$
and three different interaction strengths: $U/t=7.8$, $V/t=1.3$ (dotted line), 
$U/t=7.8, V/t=3$ (dashed line) and $U/t=7.74$, $V/t=0$ (full line). 
The broadening is $\eta/t=0.1$.
}
\label{fig:lineshapes-gfn}
\end{figure}

We now discuss the impact of varying the model parameters on the spectral 
function $A(q,\omega)$.
A comparison of the line shapes is shown in Fig.~\ref{fig:lineshapes-gfn} for 
various momenta $q$ and 
three different parameter sets ($U/t=7.8$, $V/t=1.3$), ($U/t=7.8, V/t=3$) and 
($U/t=7.74$, $V/t=0$).
These data have been calculated for a broadening  $\eta/t=0.1$ 
on a 90-site chain.
For $|q| < q_{\text F} = \pi/2$ there are only  small differences between 
the line shapes for $V/t=0$ and $V/t=1.3$.
The holon and spinon peaks shift according to the
change of the holon and spinon band widths.
This is clearly visible in the two upper panels of
Fig.~\ref{fig:lineshapes-gfn}, where spinon and holon peaks are well separated.
For a stronger coupling $V/t=3$ a significant transfer of spectral weight
occurs from the spinon peak to the holon peak. 
The low-energy continuation of the holon branch (H'')
is also visible as a small third peak in the right upper panel. 
The two middle panels of Fig.~\ref{fig:lineshapes-gfn} shows the spectral
function as  $q$ approaches $\pi/2$ from below.
There, spinon and holon structures appear to be merged in a single strong peak
because the energy difference between spinon and holon is small compared to the
broadening $\eta/t=0.1$ used.
In these panels one again sees the peak associated with the continuation
of the holon branch (H'') 
at high binding energies ($\omega/t \approx -6$). 
The spectral weight of this structure diminishes as $V$ increases and 
this structure is no longer visible for $V/t \geq 3$.

For momenta  $|q| > \pi/2$ the total spectral weight is much smaller
than for $|q| < \pi/2$, as expected, and diminishes for increasing $V$. 
Nevertheless we observe several spectral features in the
two lowest panels of Fig.~\ref{fig:lineshapes-gfn} which are usually
referred to as shadow bands.~\cite{haas} 
Shifts of the peak position for varying $U$ and $V$ are again related to the
change of the spinon- and holon band widths. 
The distribution of spectral weight is more strongly affected by the value
of the nearest-neighbor interaction for $|q| > \pi/2$ than for small 
momenta $q$. 
The peaks associated with the shadow band and the secondary holon are
well separated in the left panel ($q < \pi$) but have merged in the right
panel ($q \rightarrow \pi$). Furthermore, a low energy structure corresponding
to the lower boundary of the holon-spinon continuum is also
visible around $\omega/t=-7$. 

Finally, we have found that
there is a little spectral weight below the lower boundary of the
holon-spinon continuum. This is most clearly seen in the lowest left panel
of Fig.~\ref{fig:lineshapes-gfn} for $V/t=3$.
Therefore, excitations made of one holon and one spinon 
are not sufficient to explain all features of the one-particle 
spectral function of the extended Hubbard model.
Overall, the influence of the nearest-neighbor interaction $V$ on the 
spectral function is noticeable but not as dramatic as its impact on the
spin and charge structure factors. The main features
(spinon, holon and shadow bands) are always present with similar dispersions
albeit with different spectral weights and band widths as already found
with the variational CPT method.~\cite{potthoff}.

\section{Summary}

We have calculated the dynamical spin and charge structure factors
and
one-particle spectral function of the one-dimensional extended Hubbard model
with on-site ($U$) and nearest-neighbor ($V$) interactions using 
the dynamical density-matrix renormalization group method.
We have investigated how these dynamical correlation functions
are affected by variations of the model parameters and discussed
the spin and charge dynamics of this model in the Mott insulating regime.
For the spin structure factor the strong-coupling picture of the Heisenberg
model remains qualitatively valid down to intermediate on-site
repulsion $U \sim 4t$, in particular for the parameters relevant for cuprate
compounds. Itinerancy effects due to the coupling with charge fluctuations
become significant for weak interactions $U \alt 3t$. The nearest-neighbor
interaction $V$ does not affect the spin structure factor qualitatively.
For the charge structure factor, however, the strong-coupling approach fails
but for extremely strong interaction $U \gg t$ because the coupling to
spin fluctuations becomes relevant as soon as $t/U$ is finite.
In particular, it leads to incorrect predictions for the charge dynamics
for the parameters appropriate for cuprate compounds.
Varying the nearest-neighbor interaction $V$ causes dramatic alterations
in the charge structure factor, which reflect changes in the nature
of the charge excitations, such as the formation of (quasi-)\-excitons and CDW
droplets.
The one-particle spectral function does not change qualitatively with
varying parameters within the Mott insulating regime. 

Finally, we have shown that the extended Hubbard model allows for a
quantitative description of various experiments in the cuprate chains
SrCuO$_2$ and Sr$_2$CuO$_3$ with only a single 
choice of parameters for each material. 
For SrCuO$_2$ the extended Hubbard model with $U/t=7.8$, $V/t=1.3$ and
$t=0.435$ eV describes the low-energy optical conductivity,
the main dispersive features in the RIXS spectrum, and the 
inelastic neutron scattering spectrum quantitatively.
In addition, the  one-particle spectral function 
agrees at least qualitatively with the experimental ARPES spectrum.
For $U/t=7.4$, $V/t=1.8$, and $t=0.42$ eV the extended Hubbard model
fits the low-energy optical conductivity spectrum and the 
exchange coupling of Sr$_2$CuO$_3$ very well. 
Moreover, it reproduces the main features
of the EELS spectrum of that material at least qualitatively.
In conclusion, the extended Hubbard model yields
a coherent quantitative description of low-energy electronic
excitations in cuprate compounds
and thus provides a theoretical framework for studying the spin and charge
dynamics of these systems.

We are grateful to F.H.L.~E\ss ler, H. Frahm, F.~Gebhard
and R.M.~Noack
for helpful discussions. 
This work was supported by the Deutsche Forschungsgemeinschaft 
under Grant No NO 314/1-1.
H.B. acknowledges support by the Optodynamics Center of the
Philipps-Universit\"{a}t Marburg.
Some calculations were performed using Kazushige Goto's BLAS library.


\begin{thebibliography}{99}
%
\bibitem{dionys}
\textit{Strong Interactions in Low Dimensions}, edited by D. Baeriswyl and 
L. Degiorgi (Kluwer, 2004).
%
\bibitem{kim}
C.~Kim, A.~Y.~Matsuura, Z.~--X.~Shen, N.~Motoyama,
H.~Eisaki, S.~Uchida, T.~Tohyama, and S.~Maekawa,
Phys.~Rev.~Lett.\ \textbf{77}, 4054 (1996);
C.~Kim, Z.~-X.~Shen, N.~Motoyama, H.~Eisaki,  S.~Uchida,
T.~Tohyama, and S.~Maekawa,
Phys.~Rev.~B \textbf{56}, 15589 (1997).
%
\bibitem{koi06} 
A.~Koitzsch, S.V.~Borisenko, J.~Geck, V.B.~Zabolotnyy, M.~Knupfer, J.~Fink,
P.~Ribeiro, B.~B\"{u}chner, and R.~Follath, \prb~\textbf{73}, 201101(R) (2006).
%
\bibitem{voit} 
J.~Voit, Rep.~Prog.~Phys.~\textbf{58}, 977 (1995).
%
\bibitem{controzzi}
D.~Controzzi and F.~H.~L.~Essler,
Phys.~Rev.~B \textbf{66}, 165112 (2002).
%
\bibitem{grage}
M.J.~Bhaseen, F.~H.~L.~Essler, and A.~Grage,
Phys.~Rev.~B \textbf{71}, 020405 (2005).
%
\bibitem{penc-spectral}
J.~Favand, S.\ Haas, K.\ Penc, F.\ Mila, and E.\ Dagotto,
Phys.\ Rev.\ B \textbf{55}, R4859 (1997).
%
\bibitem{pencdensity} W.~Stephan and K.~Penc, 
Phys.~Rev.~B~{\bf 54}, R17269 (1996).
%
\bibitem{arikawa}
M.~Arikawa, Y.~Saiga, and Y.~Kuramoto,
Phys.~Rev.~Lett.\ \textbf{86}, 3096 (2001).
%
\bibitem{bannister}
R.~N.\ Bannister and N.\ d'Ambrumenil,
Phys.~Rev.~B \textbf{61}, 4651 (2000).
%
\bibitem{senechal} D.\ Senechal, D.\ Perez, 
and M.\ Pioro-\-Ladri\`{e}re, Phys.~Rev.~Lett.\ \textbf{84}, 522 (2000).
%
\bibitem{potthoff} 
M.~Aichhorn, H.G.~Evertz, W.~von~der~Linden, and M.~Potthoff,
\prb \textbf{70}, 235107 (2004).
%
\bibitem{preuss}
R.~Preuss, A.~Muramatsu, W.~von der Linden, F.~F.~Assaad, and W.~Hanke,
Phys.~Rev.~Lett.\ \textbf{73}, (1994) 732.
%
\bibitem{lavalle}
C.~Lavalle, M.~Arikawa, S.~Capponi, F.~F.~Assaad, and A.~Muramatsu,
Phys.~Rev.~Lett.\ \textbf{90}, 216401 (2003).
%
\bibitem{ddmrg}
E.~Jeckelmann,
Phys.~Rev.~B \textbf{66}, 045114 (2002).
%
\bibitem{rixs}
Y.-J.~Kim, J.~P.~Hill, H.~Benthien, F.~H.~L.~Essler, E.~Jeckelmann, 
H.~S.~Choi, T.~W.~Noh, N.~Motoyama, K.~M.~Kojima, S.~Uchida, D.~Casa, and
T.~Gog,
Phys.~Rev.~Lett.~\textbf{92}, 137402 (2004).
%
\bibitem{hubbard-spectral}
H.\ Benthien, F.\ Gebhard, and E.\ Jeckelmann,
Phys.~Rev.~Lett. \textbf{92}, 256401 (2004).
%
\bibitem{diss} {\it Dynamical Properties of Quasi One-Dimensional 
Correlated Electron Systems},
H.~Benthien,
PhD.~thesis (Marburg, 2005).
%
\bibitem{tioclpaper}
M.~Hoinkis, M.~Sing, J.~Schaefer, M.~Klemm, S.~Horn, H.~Benthien, E.~Jeckelmann,
T.~Saha-Dasgupta, 
L.~Pisani, R.~Valenti, and R.~Claessen,
Phys.~Rev.~B \textbf{72}, 125127 (2005).
%
\bibitem{matsueda}
H.\ Matsueda, N.\ Bulut, T.\ Tohyama, and S.\ Maekawa,
Phys.\ Rev.\ B \textbf{72}, 075136 (2005).
%
\bibitem{zaliznyak}
I.~A.~Zaliznyak, H.~Woo, T.G.~Perring, C.L.~Broholm, C.D.~Frost, and
H..~Takagi,
Phys.~Rev.~Lett.~\textbf{93}, 087202 (2004).
%
\bibitem{eels}
R.~Neudert, M.\ Knupfer, M.S.\ Golden, J.\ Fink, W.\ Stephan, K.\ Penc,
N.\ Motoyama, H.\ Eisaki, and S.\ Uchida, 
Phys.~Rev.~Lett.\ {\bf 81}, 657 (1998).
%
\bibitem{theBook} F.H.L.~Essler, H.~Frahm,
F.~G\"{o}hmann, A.~Kl\"{u}mper, V.E.~Korepin, \textit{The One--Di\-mensional 
Hubbard Model} (Cambridge
University Press, Cambridge, 2005).
%
\bibitem{eric-ehm}
E.~Jeckelmann,
Phys.~Rev.~Lett.~\textbf{89}, 236401 (2002).
%
\bibitem{mueller}
G.~M\"{u}ller, H.~Thomas, H.~Beck, and  J.~C.~Bonner,
Phys.~Rev.~B \textbf{24}, 1429 (1981).
%
\bibitem{fano}
U. Fano, Phys. Rev. \textbf{124}, 1866 (1961). 
%
\bibitem{moskvin} A different explanation has been proposed in
A.S. Moskvin, J.M\'{a}lek, M. Knupfer, R. Neudert, J. Fink, R. Hayn, 
S.-L. Drechsler, N. Motoyama, H. Eisaki, and S. Uchida,
\prl \textbf{91}, 037001 (2003).
%
\bibitem{eric-mott}
E.~Jeckelmann, F.\ Gebhard, and F.~H.~L.\ Essler, Phys.~Rev.~Lett.\ 
\textbf{85}, 
3910 (2000).
%
\bibitem{eric-optics-ehm}
E.~Jeckelmann,
Phys.~Rev.~B \textbf{67}, 075106 (2003). 
%
\bibitem{haas}
S. Haas and E. Dagotto, \prb \textbf{52}, 14396 (1995).

\end{thebibliography}
\end{document}